\documentclass[]{tPHM2e}

\def\qsgw{\mbox{QS$GW$}} 
\begin{document}
\doi{10.1080/1478643YYxxxxxxxx}
\issn{1478-6443}
\issnp{1478-6435}
\jvol{00} \jnum{00} \jyear{2008} \jmonth{21 December}

\markboth{}{Philosophical Magazine}


\title{GW correlation effects on plutonium quasiparticle energies: changes in crystal-field splitting}

\author{A. N. Chantis$^{\rm a}$$^{\ast}$\thanks{$^\ast$Corresponding author. Email: achantis@lanl.gov\vspace{6pt}} 
; $^{\rm a}$R. C. Albers; $^{\rm b}$A. Svane; $^{\rm b}$N. E. Christensen\\\vspace{6pt}  
$^{\rm a}${\em{Theoretical Division, Los Alamos National Laboratory,
Los Alamos, New Mexico, 87545, USA}}; $^{\rm b}${\em{Department of Physics and Astronomy, University of Aarhus, DK-8000 Aarhus C, Denmark}}
\\\vspace{6pt}\received{} }

\maketitle

\begin{abstract}
We present results for the electronic structure of plutonium by
using a recently developed quasiparticle self-consistent $GW$ method (\qsgw). 
We consider a paramagnetic solution without spin-orbit interaction as a function of
volume for the face-centered cubic (fcc) unit cell. We span unit-cell volumes ranging from
10\% greater than the equilibrium volume of the $\delta$ phase to 90 \% of
the equivalent for the $\alpha$ phase of Pu. The self-consistent $GW$ quasiparticle energies
are compared to those obtained within the Local Density Approximation (LDA).
The goal of the calculations is to understand systematic trends 
in the effects of electronic correlations on the quasiparticle
energy bands of Pu as a function of the localization of the $f$ orbitals.  
We show that correlation effects narrow the $f$ bands in two significantly
different ways.  Besides the expected narrowing of individual $f$ bands
(flatter dispersion), we find that an even more significant effect on the
$f$ bands is a decrease in the crystal-field splitting of the different bands.
 
\bigskip

\end{abstract}

\section{Introduction}

Much of our modern understanding of electronic correlations in narrow-band systems
has derived from many-body treatments of model-Hamiltonian systems such as
the Hubbard model and the Anderson model.  For example, with respect to plutonium
in particular, dynamical mean-field theory (DMFT) approaches have been very useful
in elucidating the physics of the very strong correlations in this material (see, for example,
Refs.~\cite{dmft1} and \cite{dmft2} and references therein).  Nonetheless, these calculations
are not first principles, and most of the physics comes from the model part of the Hamiltonian
rather than the band-structure part of the calculations.
Thus, it is still important to better understand the multi-orbital and hybridization 
effects in more realistic electronic-structure approaches that are less model dependent.  
The GW method is our best modern tool to examine these effects,
because it includes correlation effects beyond that of conventional local-density approximation
(LDA) band-structure techniques and yet is still first-principles.

In this paper we study correlation effects of fcc Pu as a function of volume. Our goal is not to
specifically elucidate the correlation physics of Pu itself, since the GW approach is a low-order approximation
and cannot treat the very strong correlations of the $\delta$ phase of Pu.  Rather, we wish to use
the volume dependence to tune the material from high atomic density (high pressure), where correlations can
be greatly reduced due to the large hybridization between the Pu $f$ orbitals, to low atomic densities (where the
pressure would actually be in tension), where the correlations effects are very strong.  By following this
procedure, we can understand how correlations modify the properties of a material
$within\ the\ GW\ approximation$.  When the correlations become strong, it is certainly the case that higher-order
approximations like DMFT are necessary to accurately describe the material.  Nonetheless, since GW is
probably the correct starting point for such types of more sophisticated approaches, it is still useful to understand what
happens to the electronic-structure of the material within the GW approximation as a function of the strength
of the correlation effects, and, in particular, their effects on shifts in quasi-particle energies, which is what GW is
best at representing.

In order to
focus more specifically on the effects of correlations, we ignore one significant
aspect of the electronic structure, viz., the spin-orbit coupling, even though this is important
for a detailed comparison with experiment.
Spin-orbit coupling mainly shifts $f$ states around in energy, but at the same time add to the complexity of the
individual $f$ bands, hence tending to hide some of the correlation physics in a bewildering array
of bands.  Spin-orbit effects
can easily be added in when a more accurate comparison with experiment is desired (as was done for
an earlier paper on less correlated $\alpha$ uranium \cite{chantis2}).

For the same reasons we also ignore well-known large changes in crystal structure of Pu metal with volume, and present
calculations only for the simple fcc crystal structure as a function of volume.  We study a range of Pu atomic volumes
extending from well below that pertinent to the ground state $\alpha$ phase to well above that
of the high-temperature $\delta$ phase.
The actual crystal structure of $\alpha$Pu is  a complicated monoclinic structure with 16 atoms per unit cell,
while  $\delta$Pu has the fcc structure.
Since $\delta$Pu is well known to be a strongly correlated-electron metal while $\alpha$Pu appears
to be reasonably well treated by conventional band-structure methods, we believe that our range
of volumes corresponds to tuning the correlation effects between weak to moderate (small volumes) to
strongly correlated (large volumes).

In the calculations to be presented in the following we mainly focus on changes in the effective bandwidth of the $f$ states in Pu.
We will show that crystal-field effects are actually a more important factor in determining this bandwidth
than the expected change in dispersion (flattening of the bands).

\section{Method}

The $GW$ approximation can be viewed as the first term in the
expansion of the non-local energy-dependent self-energy
$\Sigma(\bf{r},\bf{r}',\omega)$ 
in the screened Coulomb interaction
$W$. From a more physical point of view it can be interpreted as a
dynamically screened Hartree-Fock approximation plus a Coulomb-hole
contribution \cite{Hedin}.  It is also a prescription for mapping the
non-interacting Green function onto the dressed Green's function: $G^0 \to G$.  This
prescription can be described as follows. From the Hamiltonian 
\begin{equation}
H_0 = {-\bigtriangledown^2} + V_{\rm{eff}}(\mathbf{r},\mathbf{r^\prime})
\end{equation}
(we use atomic Rydberg units: $\hbar=2m=e^2/2=1$, where $m$ and $e$ are the mass and charge of the electron) 
$G^0=(\omega-H_0)^{-1}$ may be constructed. 
Often $G^0$ is calculated from the
LDA eigenvalues and eigenfunctions; however, there is no formal
restriction for how to choose the initial starting point $G^0$.  Then,
using the Random Phase Approximation (RPA) \cite{Hedin}, we can construct
the polarization function $D$ and screened Coulomb interaction $W$ as
\begin{equation}
D = -i G^0 \times G^0
\label{PolFct}
\end{equation}
and
\begin{equation}
W = u \times \frac{1}{1-uD} ~,
\end{equation}
where $u$ is the bare Coulomb interaction.  The new Green's function
is defined as
\begin{equation}
G = \frac{1}{\omega - ( {-\bigtriangledown^2}+ V^{\rm{ext}}+ V^{\rm{H}}+\Sigma)}~,
\end{equation}
where $V^{\rm{ext}}$ is the potential due to ions (Madelung) and $V^{\rm{H}}$ is the Hartree
potential
\begin{equation}
V^{\rm{H}}(\mathbf{r}) = 2 \int d \mathbf{r^{\prime}} \frac{n(\mathbf{r^\prime})}{|\mathbf{r}-\mathbf{r^\prime}|} ~.
\end{equation}
The single-particle density $n(\mathbf{r})$ is calculated as
$n(\mathbf{r})\sim\int_{-\infty}^{\infty}d\omega
G^0(\mathbf{r},\mathbf{r},\omega)e^{i\omega\delta}$. 

In addition to
this mapping of $G^0 \to G$, one can also generate an excellent effective potential
from $G$ that makes it possible to approximately do the inverse mapping of $G \to
G^0$ \cite{Mark2}.  The \qsgw\ is a method to specify this (nearly)
optimal mapping of $G \to G^0$, so that $G^0 \to G \to G^0 \to ...$ can
be iterated to self-consistency \cite{Mark2,kotani1}.  At self-consistency the
quasiparticle energies of $G^0$ coincide with those of $G$.  Thus
\qsgw\ is a self-consistent perturbation theory, where the
self-consistency condition is constructed to minimize the size of the
perturbation.  The \qsgw\ method is parameter-free, and independent of
basis set as well as the LDA starting point \cite{Mark2,kotani1,Mark1}.  We have
previously shown that \qsgw\ reliably describes a wide range of
$spd$~\cite{Mark2,Faleev,chantis1,chantis4,kotani2}, and rare-earth~\cite{chantis3} systems.
We have also applied the method to calculate the electronic structure 
of $\alpha$-uranium \cite{chantis2}.

Our version of the \qsgw\ method is based upon the Full Potential
Linear Muffin Tin Orbital (FP-LMTO) method \cite{markbook}, which
makes no approximation for the shape of the crystal potential.
The smoothed LMTO basis
\cite{Mark1} includes orbitals with $l \le l_{max}=6$; both 7$p$
and 6$p$ as well as both 5$f$ and 6$f$ are included in the basis. The
$6f$ orbitals are added as a local orbital \cite{Mark1}, which is
confined to the augmentation sphere and has no envelope function.  The
$7p$ orbital is added as a kind of extended `local orbital,' the
`head' of which is evaluated at an energy far above Fermi level
\cite{Mark1} and instead of making the orbital vanish at the
augmentation radius a smooth Hankel `tail' is attached to the orbital.
The  7$p$ and 6$f$ orbitals are necessary to obtain an accurate description of highlying bands,
which are important for the accuracy of the polarization function in Eq. (\ref{PolFct}).
For our calculations we use the fcc Pu lattice with the following
lattice constants $a$=4.11 $\AA$ (which corresponds to 90$\%$ of the $\alpha$-Pu equilibrium volume), 
4.26 $\AA$ (at the $\alpha$-Pu equilibrium volume), 4.64 $\AA$ (at the $\delta$-Pu equilibrium volume)
and 4.79 $\AA$ (at 110$\%$ of the $\delta$-Pu equilibrium volume).

\section{Results}

\begin{figure}[htbp]
\includegraphics[angle=0,width=0.45\textwidth,clip]{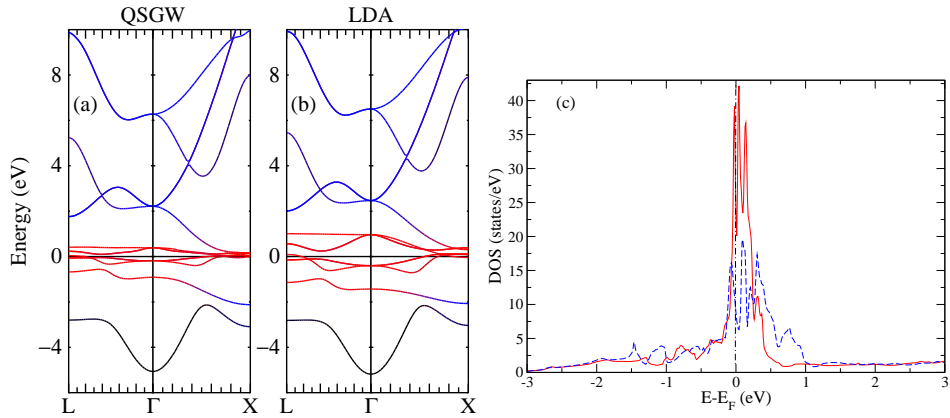}
\includegraphics[angle=0,width=0.45\textwidth,clip]{Fig2.eps}
\caption{ (color online).  (a) The QSGW energy bands (or quasi-particle energies)
for $\delta$-Pu along two
symmetry directions (left panel), compared to (b) the LDA energy bands (right panel);
the Mulliken weights of the $f$ orbitals are presented in red (dark gray), for $s$ orbitals in black
and for $d$ orbitals in  blue (light gray). (c)  Comparison of the total density of states (DOS)
for QSGW, red (dark gray) 
solid line, and LDA, blue (light gray) dashed line. The Fermi energy is set at zero.}
\label{fig1}
\end{figure}

In Fig.1 we compare the QSGW one-particle electronic structure of
$\delta$-Pu with the LDA band-structure results. The spin-orbit
interaction is not included in this calculation. The Mulliken weights 
of the $f$ orbitals are presented in red (dark gray), of the $s$ orbitals in black
and of the $d$ orbitals in  blue (light gray). In both cases, the narrow bands
located between -2 and 2 eV are predominantly due to seven 5$f$ orbitals.
At the $\Gamma$ point they are split by the cubic crystal field into 
one nondegenerate and two three-fold degenerate states. This degeneracy 
is reduced at general k-points in the Brillouin zone.
The lowest dispersive band centered around -4 eV has primarily $s$ character and
the unoccupied bands above 2 eV are mainly due to $d$ orbitals.
At the $\Gamma$ point the five $d$ states are split by the cubic crystal field into
one two-fold degenerate and one three-fold degenerate state. 
The $sdf$ hybridization along the symmetry directions presented in 
Fig.~\ref{fig1} is generally very weak except for near the $X$ point along the
$\Gamma-X$ direction, where there is very strong $sd$ and $df$
hybridization for some of the $d$ and $f$ branches.
The degree of the hybridization is very similar in both
the LDA and $QSGW$ calculations, as is
the center of all of the bands. The largest visible change is
a significant narrowing of the $5f$-band complex. This effect has two components:
first, the crystal-field splitting of the $f$ bands is significantly reduced
in \qsgw, second, the bandwidth of each individual $f$ branch (band) 
is also reduced in \qsgw.
The total effect appears as an overall narrowing of the total 
density of states (DOS) around the Fermi level (see Fig.~\ref{fig1}(c)).
In addition, since the area under the curve is proportional to
the number of $5f$ states, which remains constant, the 
amplitude of the quasiparticle peaks are also higher in QSGW.
For example, the total DOS at the Fermi level in QSGW is 23 states/eV while in LDA
it is 9 states/eV.

\begin{figure}[bp]
\centering
\includegraphics[angle=0,width=0.55\textwidth,clip]{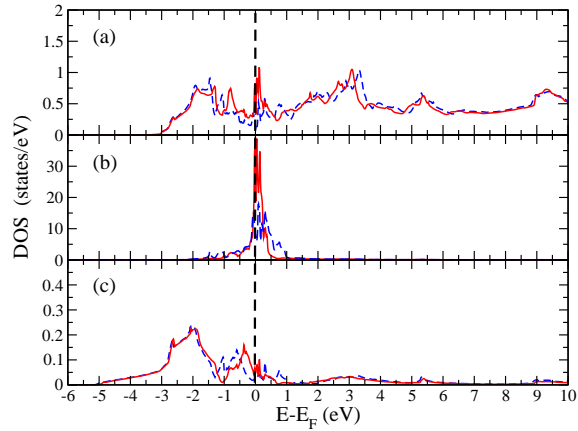}
\caption{ (color online). Comparison of QSGW, red (dark gray) solid line, and LDA, blue (light gray) dashed line, 
partial DOS for (a) the $d$ orbitals, (b) the $f$ orbitals, and (c) the $s$ orbitals.}
\label{fig2}
\end{figure}

The partial DOS is presented in Fig.~\ref{fig2}. In both calculations the partial-$5f$ DOS is 
concentrated in a narrow energy interval around $E_F$. The partial-$s$ DOS is mainly located in the
occupied energy spectrum between -5 eV and $E_F$ and the $d$ bands are spread in 
a wide energy interval in both the unoccupied and occupied part of the spectrum with a few 
pronounced peaks at various energies. 
Overall, the \qsgw\ and LDA $s$ and $d$ peaks are located at the same energies, 
with the exception of the narrow peaks around $E_F$, 
in which case the \qsgw\ are visibly shifted closer to $E_F$. 
In this region these bands can be highly hybridized with the $f$ states, 
and thus these shifts reflect the band narrowing of the $f$ states. 
The bottom of the $s$ and $d$ bands relative to the position of $f$ band is 
approximately at the same energy location in \qsgw\ and LDA. 
The $f$ occupation changes in \qsgw\ from its
LDA value of 5.06 to 4.85 states. A similar reduction was observed in our calculations
for $\alpha$-uranium~\cite{chantis2}.

\begin{figure*}[tbp]
\includegraphics[angle=0,width=0.5\textwidth,clip]{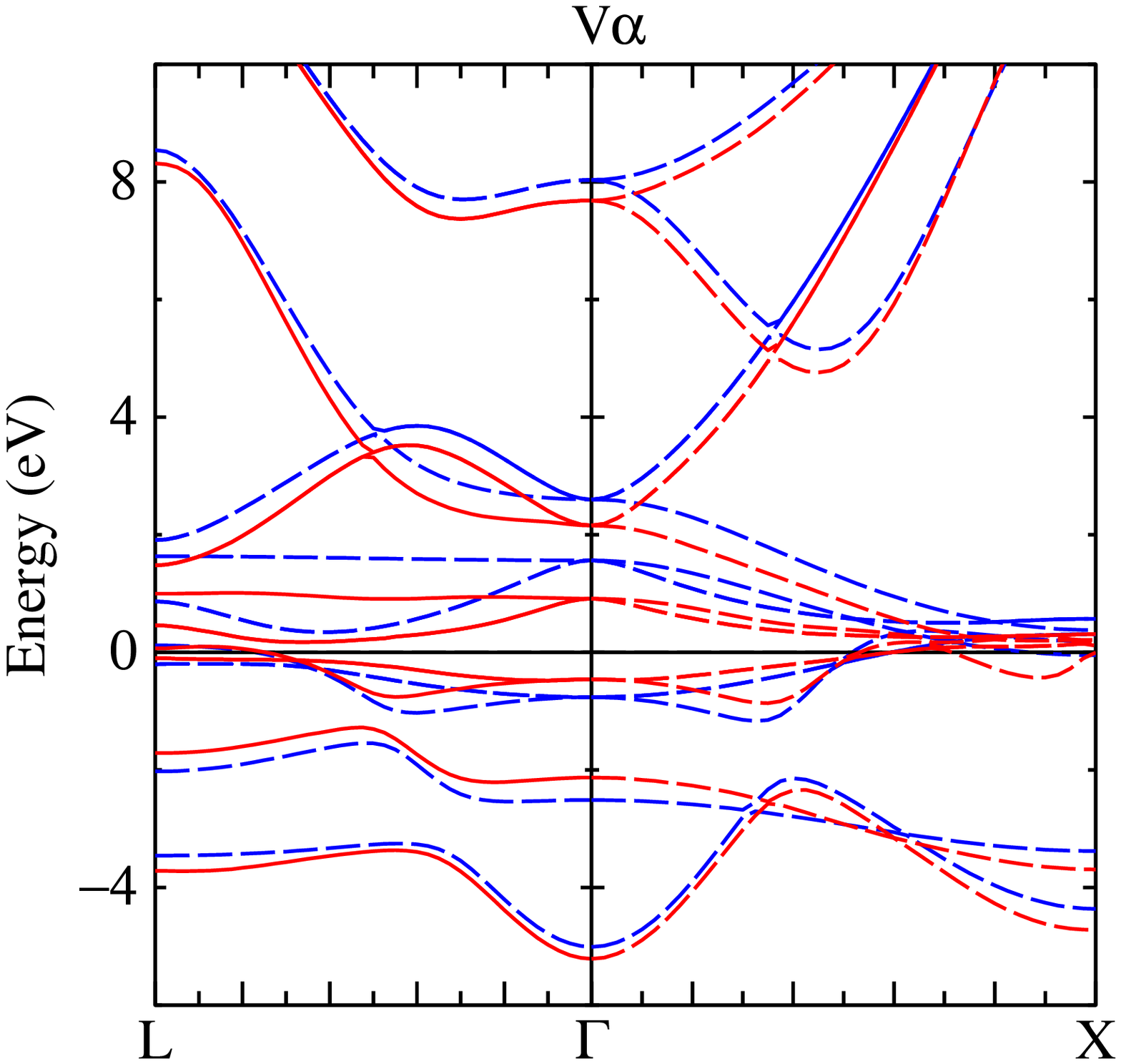}
\includegraphics[angle=0,width=0.5\textwidth,clip]{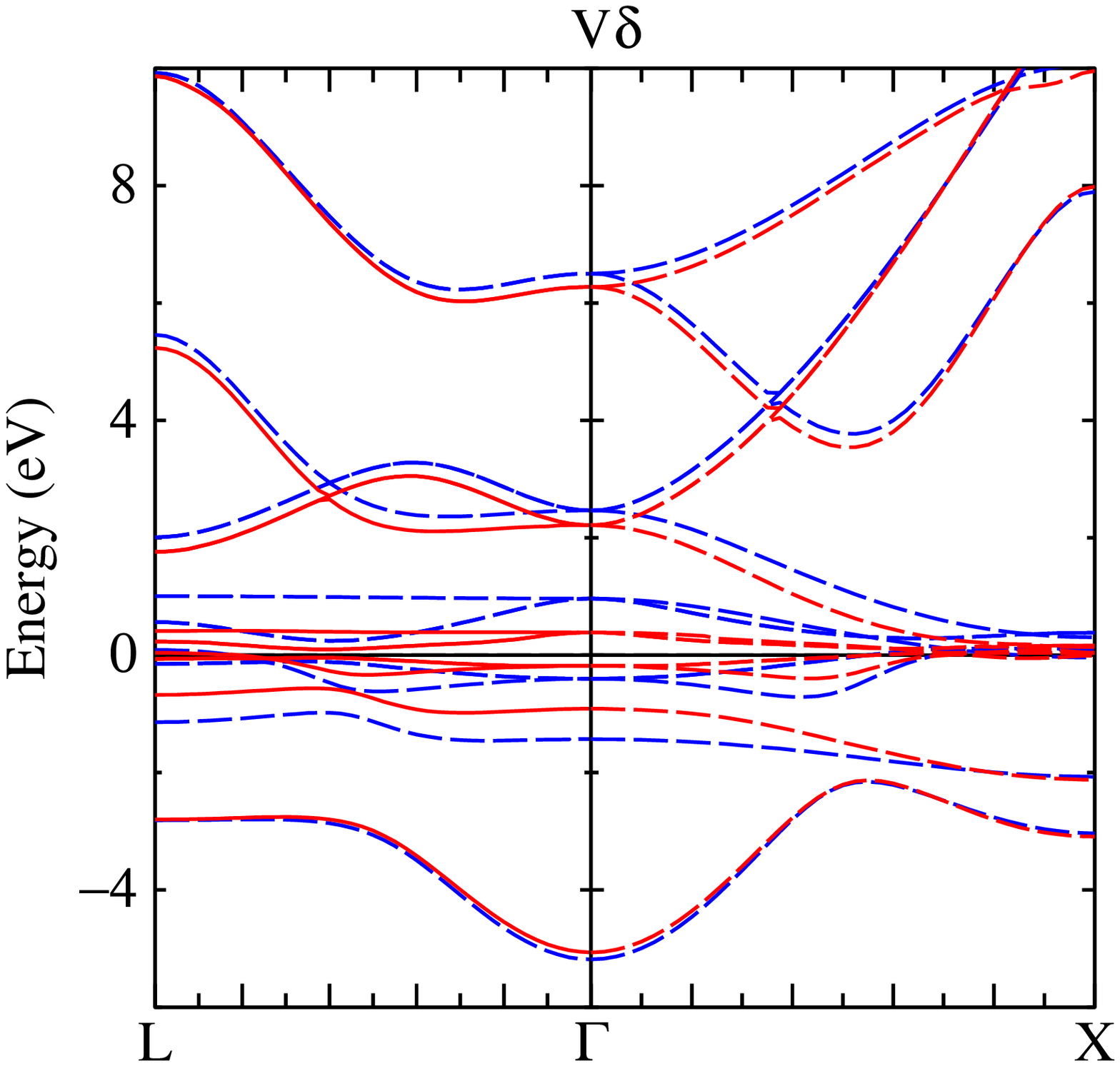}
\includegraphics[angle=0,width=0.5\textwidth,clip]{pdos-3.eps}
\includegraphics[angle=0,width=0.5\textwidth,clip]{pdos-4.eps}
\caption{ (color online). Comparison of the QSGW, red (dark gray) solid line, and LDA, blue (light gray) dashed line, 
band structure. On the left side are the energy bands for the more weakly
correlated case for $a$=4.26 $\AA$ (equivalent to the density of atoms for the $\alpha$-Pu equilibrium volume), 
and on the right side the more strongly correlated case of $a$=4.64 $\AA$ (the $\delta$-Pu equilibrium volume).}
\label{fig3}
\end{figure*}

In Fig.~\ref{fig3} we present side by side the band structures for two different fcc volumes.
On the left side is the band structure and partial $s$ and $d$ DOS for the
$\alpha$ volume (weakly correlated) and on the right side is the band structure
and partial $s$ and $d$ DOS for the $\delta$ volume (strongly correlated).
In all cases, the red solid lines represent the \qsgw\ results and the dashed-blue lines
show the LDA results.
Two major effects are seen in this plot: (1) the $f$ bands narrow considerably with expanded
volume, and (2) there is a similar effect on the $d$ bands (notice the downward shift of
the bands at the highest energy as one goes from the $\alpha$ to the $\delta$ volume).
   
\begin{figure}[tbp]
\centering
\includegraphics[angle=0,width=0.55\textwidth,clip]{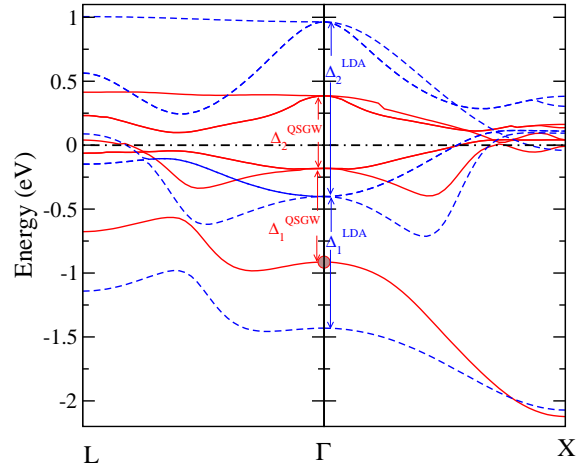}
\caption{ (color online). Comparison of QSGW, red (dark gray) solid line, and LDA,
blue (light gray) dashed line, energy bands along two symmetry directions.
The magnitude of crystal field splittings $\Delta_1$, $\Delta_2$ are
1.02, 1.37 eV in LDA and 0.73, 0.57 eV in QSGW. $\Delta_2$ is
reduced significantly in QSGW calculation. For the 5$f$ state at the $\Gamma$ point, marked with a large dot, in Fig.~\ref{fig4b} we present the energy dependence of the self-energy and calculated spectral function $A(\omega)$}
\label{fig4}
\end{figure}

\begin{figure}[bp]
\centering
\includegraphics[angle=0,width=0.55\textwidth,clip]{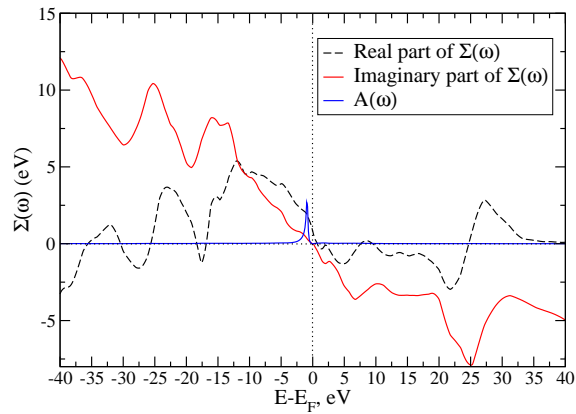}
\caption{The energy dependence of the real and imaginary part of self-energy together with the
spectral function for the quasiparticle at ${\bf k}=(0,0,0)$ and $E_0=-0.917$ eV.}
\label{fig4b}
\end{figure}
\begin{figure}[bp]
\centering
\includegraphics[angle=0,width=0.55\textwidth,clip]{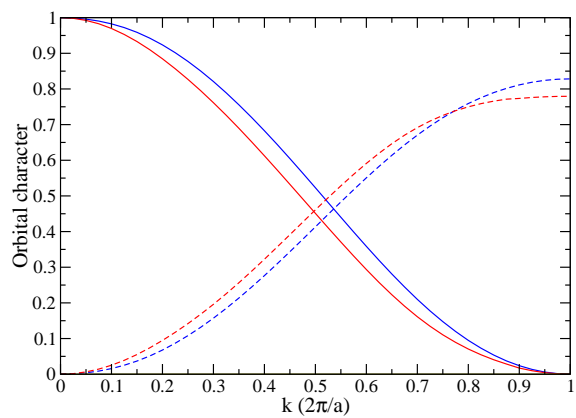}
\caption{ (color online). QSGW, red (dark gray) solid line, and LDA, blue (light gray)
solid line $f$-orbital weight of the $A_{2}$ band
along $\Gamma-X$ direction. Also shown is the QSGW, red (dark gray) dashed line,
and LDA, blue (light gray) dashed line $d$-orbital weight 
of the $A_{2}$ band along the same direction}
\label{fig5}
\end{figure}

     
To examine band narrowing effects in more detail, in Fig.~\ref{fig4} we expand the view of the 5$f$ bands.
Also, in Fig.~\ref{fig4b} we show the self-energy and spectral function for one of the \qsgw  5$f$ states.
Despite of the complicated energy dependence of the real and imaginary parts it appears that the state
is described perfectly well by Landau's Fermi liquid theory. The imaginary part of the self-energy
goes through zero at $E_F$ and the spectral function has a \emph{single} 
well defined peak centered at the \qsgw  eigenvalue. There are no other pronounced features in the energy
dependence of the spectral function. This is representative of all $5f$ states. 
So the 5$f$ states within the \qsgw  theory are well defined quasiparticles with very large 
lifetime around $E_F$; the \qsgw  eigenvalues coincide with the quasiparticle energy.
Therefore, our discussion is focused on the effects of electron correlations on the \qsgw  eigenvalues.
At the $\Gamma$ point the $f$ orbitals are split by the cubic crystal field into a
nondegenerate state $A_{2}$ and two three fold degenerate states $T_{1}$ and $T_{2}$.
The splitting, $\Delta_1$, between the nondegenerate state and the lowest
of the three-fold degenerate states
is 1.02 eV in LDA and 0.73 eV in \qsgw.
The splitting, $\Delta_2$, between the two three-fold degenerate states is equal to 1.37 eV in LDA
but only 0.57 eV in \qsgw. 
This significant reduction of the crystal field splitting in \qsgw\ is 
the major part of the band narrowing observed in the DOS in Fig.~\ref{fig1}. 
Another important aspect is the reduction of
the width of each individual $f$ band. In Table~\ref{table1} we present the
values of the band width of Pu $5f$ bands along $L-\Gamma$ and $\Gamma-X$ symmetry
directions. The band width is defined as the difference between the
maximum and the minimum energy of a particular band along the direction.  
In all cases the band width is reduced significantly in \qsgw. 
The $A_{2}$ band along $\Gamma-X$ is a striking exception from this rule.
As we show in Fig.~\ref{fig5} this band strongly hybridizes with the $d$ states.
At the $\Gamma$ point it is 100$\%$ $f$ but at the $X$ point it is predominately
$d$. All the other bands shown in Fig.~\ref{fig4} remain 80 to 100$\%$ $f$ throughout
the symmetry directions shown in the figure. This explains the anomalous change 
in the width of this 
band for this particular direction. From Fig.~\ref{fig3} it is evident that,
while in \qsgw\ the width of the $f$ bands are significantly reduced, the width 
of $d$ bands are practically the same as for LDA. The $A_{2}$ band
at the $\Gamma$ point has mainly $f$ character and therefore moves upward
from its LDA position due to the significant reduction of the $f$ crystal field in \qsgw, 
but at the
$X$ point mainly has $d$ character and therefore remains approximately 
at its LDA position (only slightly lower due to the slight downward shift of the
center of $d$ band in \qsgw). The cumulative effect is that in \qsgw\ this band
is stretched. 

\begin{table}
\tbl{The band width of Pu $5f$ bands along $L-\Gamma$ and $\Gamma-X$ symmetry
directions. The band width is defined as the difference between the
maximum and the minimum energy of a particular band along the direction.
The width is given in eV}
{\begin{tabular}{lcc}\toprule
                               &$L-\Gamma$&  $\Gamma-X$\\
\colrule
$A_{2}$ LDA        & 0.48  & 0.64  \\
$A_{2}$ QSGW       & 0.43  & 1.21  \\
\colrule
$T_{1}$ LDA   & 0.71   & 0.83 \\
$T_{1}$ QSGW   & 0.38  & 0.44 \\
\colrule
$T_{2}$ LDA  & 0.77 & 1.02 \\
$T_{2}$ QSGW  & 0.32 & 0.45 \\
\botrule
\end{tabular}}
\label{table1}
\end{table}

\begin{table}
\tbl{The crystal field splitting $\Delta_{1}$ and $\Delta_{2}$ of Pu 5f states at the
$\Gamma$ point. 
The $QSGW$ values for the splitting are significantly smaller than those of LDA calculation. 
The energy splitting is given in eV}
{\begin{tabular}{lcccc}\toprule
                               &0.9 $V_{\alpha}$&  $V_{\alpha}$ & $V_{\delta}$& 1.1 $V_{\delta}$      \\
\colrule
$\Delta_{1}^{LDA}$        & 2.14  & 1.75  & 1.02 & 0.84\\
$\Delta_{1}^{QSGW}$       & 2.09  & 1.67  & 0.73 & 0.51\\
\colrule
$\Delta_{2}^{LDA}$         & 2.86   & 2.33  & 1.37  &  1.11  \\
$\Delta_{2}^{QSGW}$ & 1.96 & 1.36 & 0.57  &   0.41  \\
\botrule
\end{tabular}}
\label{table2}
\end{table}

\begin{table}
\tbl{The band width of Pu $5f$ bands along $L-\Gamma$ symmetry
directions for different volumes of the unit cell. The band width is defined as 
the difference between the
maximum and the minimum energy of a particular band along the direction. 
We also provide the full width at half maximum (FWHM) of the 
broadened $f$-partial DOS shown in Fig.~\ref{pdos}. 
The width is given in eV}
{\begin{tabular}{lcccc}\toprule
                             &0.9 $V_{\alpha}$&  $V_{\alpha}$ & $V_{\delta}$& 1.1 $V_{\delta}$  \\
\colrule
$A_{2}$ LDA          & 1.42  & 1.0  & 0.48 & 0.37 \\
$A_{2}$ QSGW         &  1.23 & 0.94 &  0.43&  0.30\\
\colrule
$T_{1}$ LDA  & 1.38 & 1.15 & 0.71 & 0.6  \\
$T_{1}$ QSGW &  1.17 & 0.86 &  0.38&  0.28\\
\colrule
$T_{2}$ LDA  & 1.62  & 1.29 & 0.77 & 0.63\\
$T_{2}$ QSGW &  1.30 & 0.84 &  0.32&  0.23\\
\colrule
FWHM LDA  & 1.28  & 1.09  & 0.78 & 0.72\\
FWHM QSGW & 0.79  & 0.63 & 0.39 & 0.34 \\
\botrule
\end{tabular}}
\label{table3}
\end{table}

\begin{figure}[tbp]
\centering
\includegraphics[angle=0,width=0.65\textwidth,clip]{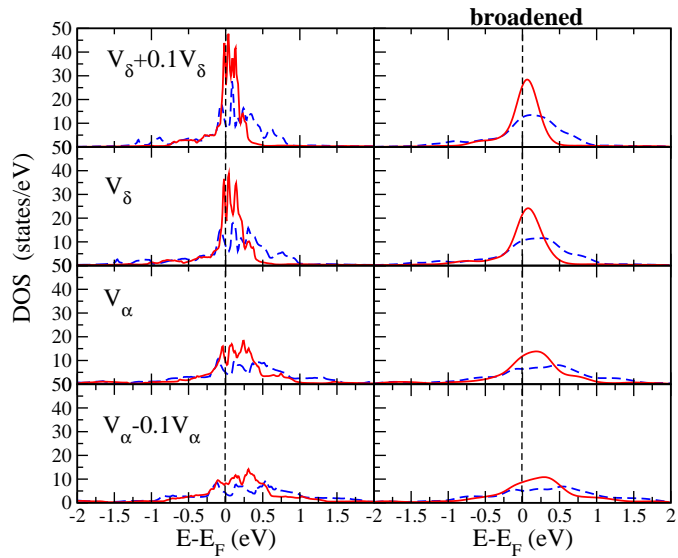}
\caption{ (color online). The QSGW and LDA $f$-partial DOS for four different volumes. 
The panels on the right side show the corresponding
DOS on the left panel broadened with a Gauss function.}
\label{pdos}
\end{figure}

In Tables \ref{table2} and \ref{table3} we show the values for the crystal field splitting
of 5$f$ bands and $f$ band widths along $L-\Gamma$ for several volumes of the unit cell.  
The band width and crystal field splitting of all bands is reduced as we move from 
the lower to higher volume case. This is a result of the reduction of the 
$f$ band relative extend in the crystal. It is also seen that the \qsgw\ 
crystal field splittings and band widths for $f$ orbitals are always smaller than 
in LDA. We have also considered the band width of the entire 5$f$ band complex. 
To do this we have applied a Gaussian broadening on the partial-$f$ DOS (Fig.~\ref{pdos}).
In this case the  5$f$ band complex appears like a single large peak. We can define
the width of this band as the full width at half maximum (FWHM)
of the peak. This is also presented in Table~\ref{table3}. It is evident that 
in both calculations the width is reduced as the volume increases.
The rate of reduction is the same in both calculations.    
But the \qsgw\ FWHM is always significantly smaller than the LDA.    
Therefore, we conclude that in \qsgw\ the $f$ orbitals contract due to the
more accurate treatment of correlations. On the other hand, the $s$ and $d$
bands are very itinerant and are therefore already described accurately at the
level of the LDA treatment of correlations.

\section{Conclusions}

In conclusion, we have applied \qsgw\ theory to $\delta$ plutonium as a function
of volume in order to systematically understand the effects of electronic correlation
on the band narrowing of the $f$ bands.
Unlike conventional model-Hamiltonian treatments of 
strongly correlated systems our approach is first principles and independent of
any choice of model parameters, and hence
provides a unique opportunity to examine the effect of electron correlations on the
quasiparticle band structure as a function of $f$-orbital localization.
In this way we have demonstrated that \qsgw\ and LDA prediction for the $s$ and $d$
electron subsystems are quite similar. This is because these electrons are very
itinerant and therefore their description lies within the validity of LDA.
However, the \qsgw\ $5f$ bands are much narrower than their LDA counterpart.
We believe that our results for the first time show significant details of the
band narrowing due to electron correlations that have not been previously studied. In
particular, the \qsgw\ calculations show that the major contribution to band
narrowing is actually the reduction of
the crystal-field splitting of 5$f$ states as compared
to effects from a reduction in dispersion (flattening of the individual bands). 
This is a significant change in the character of the 5$f$ states and suggests
the importance of using GW approaches as input to more sophisticated correlation
approaches such as dynamical mean-field theories (DMFT).   

\section{Acknowledgments}
This work was carried out under the auspices of the National Nuclear Security
Administration of the U.S. Department of Energy at Los Alamos National Laboratory 
under Contract No. DE-AC52-06NA25396. A.N.C would like to thank 
Mark van Schilfgaarde and Takao Kotani for informative discussions.

\end{document}